\documentclass[prl,showpacs,showkeys,nofootinbib,twocolumn]{revtex4}

\usepackage{graphicx}  %
\usepackage{bm}  %

\newcommand{\boma}[1]{\mbox{\boldmath{$#1$}}}

\newcommand{\beq}{\begin{equation}}
\newcommand{\eeq}{\end{equation}}
\newcommand{\bqa}{\begin{eqnarray}}
\newcommand{\eqa}{\end{eqnarray}}

\def\mqo2{{\!\!\!}}

\begin{document}

\title{An Infrared Renormalization Group Limit Cycle in QCD}

\author{Eric Braaten}
\affiliation{Department of Physics,
         The Ohio State University, Columbus, OH\ 43210, USA}

\author{H.-W. Hammer}
\affiliation{Helmholtz-Institut f{\"u}r Strahlen- und Kernphysik 
   (Abt.~Theorie), Universit{\"a}t Bonn, 53115 Bonn, Germany}

\date{March 17, 2003}

\begin{abstract}
Small increases in the up and down quark masses of QCD
would tune the theory to the critical renormalization group 
trajectory for an infrared limit cycle 
in the three-nucleon system.
At critical values of the quark masses,
the deuteron binding energy goes to zero and
the triton has infinitely many excited states 
with an accumulation point at the 3-nucleon threshold.
The ratio of the binding energies of successive states 
approaches a universal constant that is close to 515.
The proximity of physical QCD to the critical trajectory 
for this limit cycle explains the success of an effective field theory 
of nucleons with contact interactions only in describing
the low-energy 3-nucleon system. 
\end{abstract}

\smallskip
\maketitle

The development of the renormalization group (RG)
has had a profound effect on many branches of physics.
Its successes range from explaining the universality
of critical phenomena in condensed matter physics
to the nonperturbative formulation of quantum field theories
that describe elementary particles \cite{Wilson:dy}.  
Most simple applications of the renormalization group involve 
fixed points that are approached either in the infrared limit 
or in the ultraviolet limit.
However, the renormalization group can be reduced to a set of 
differential equations that define a flow in the space of 
coupling constants.  A fixed point where the flow vanishes
is only the simplest topological feature that can be 
exhibited by such a flow.  
Another topological feature is a limit cycle, 
which is a 1-dimensional orbit that is closed under the RG flow.
The possibility of RG flow to a limit cycle was proposed by 
Wilson in 1971 \cite{Wilson:1970ag}.  
Glazek and Wilson have recently
constructed a simple discrete Hamiltonian system that exhibits
RG flow to a limit cycle \cite{Glazek:2002hq}.
However, few physical applications of RG limit cycles have emerged.

The purpose of this letter is to point out that Quantum Chromodynamics
(QCD) is close to the critical trajectory for an infrared RG limit cycle
in the 3-nucleon sector.  
It can be tuned to the critical trajectory by small changes 
in the up and down quark masses.  
The proximity of the physical quark masses to these critical values
explains the success of a program initiated by Efimov
to describe the 3-nucleon problem in terms of zero-range forces
between nucleons \cite{Efi81}.
An effective-field-theory formulation of this program by 
Bedaque, Hammer, and van Kolck
exhibits an ultraviolet RG limit cycle \cite{Bedaque:1999ve}.
The proximity of physical QCD to the critical trajectory for an
infrared limit cycle shows that the ultraviolet limit cycle of 
Ref.~\cite{Bedaque:1999ve}
is not just an artifact of their model.

In the late 1960's, Wilson used the renormalization group (RG)
to explain universality in critical phenomena \cite{Wilson:dy}.
Transformations of a system that integrate out 
short-distance degrees of freedom while leaving
the long-distances physics invariant define a 
{\it RG flow} on the multidimensional space 
of coupling constants ${\bf g}$ for all possible terms in the Hamiltonian:
\begin{eqnarray}
\Lambda {d \ \over d \Lambda} {\bf g} = \boma{\beta}({\bf g} ) \,,
\label{RGEq}
\end{eqnarray}
where $\Lambda$ is an ultraviolet momentum cutoff.
Standard critical phenomena are associated with 
{\it infrared fixed points} ${\bf g}_*$ of the RG flow,
which satisfy $\boma{\beta}({\bf g}_*) = 0$.
The tuning of macroscopic variables to reach a critical point 
corresponds to the tuning of the coupling constants ${\bf g}$
to a {\it critical trajectory} that flows to the fixed point ${\bf g}_*$
in the infrared limit $\Lambda \to 0$. 
One of the signatures of an RG fixed point
is {\it scale invariance}: symmetry with respect to the 
coordinate transformation ${\bf r} \to \lambda {\bf r}$ 
for any positive number $\lambda$.  
This symmetry implies that dimensionless variables 
scale as powers of the momentum scale, perhaps with anomalous dimensions.

In a classic 1971 paper, Wilson suggested that 
the renormalization group might also be relevant to the strong 
interactions of elementary particle physics \cite{Wilson:1970ag}.
At that time, the fundamental theory for the strong interactions
had not yet been discovered.  It was believed 
to involve quarks, and hints that the strong interactions might have
scaling behavior at high energies had been observed in experiments 
on deeply-inelastic lepton-nucleon scattering.
Wilson suggested that simple high-energy behavior can be explained
by simple RG flow of the relevant coupling constants 
in the ultraviolet limit $\Lambda \to \infty$.  The simplest possibility 
is RG flow to an {\it ultraviolet fixed point}.
Another simple possibility is RG flow to an {\it ultraviolet limit cycle}.
A limit cycle is a 1-parameter family of coupling constants 
${\bf g}_*(\theta)$ that is closed under the RG flow
and can be parametrized by an angle $0 < \theta < 2 \pi$.
The RG flow carries the system 
around a complete orbit of the limit cycle every time
the ultraviolet cutoff $\Lambda$ increases by some factor $\lambda_0$.
One of the signatures of an RG limit cycle
is {\it discrete scale invariance}: symmetry with respect to the 
coordinate transformation ${\bf r} \to \lambda_0^n {\bf r}$ 
only for integer values of $n$.  
This symmetry implies that dimensionless variables 
with vanishing anomalous dimensions are periodic functions of
the logarithm of the momentum scale with period $\ln(\lambda_0)$.
The fundamental field theory 
for the strong interactions, QCD, was eventually discovered.  
QCD has a single coupling constant $\alpha_s(\Lambda)$ 
with an {\it asymptotically-free} ultraviolet fixed point: 
$\alpha_s(\Lambda) \to 0$ as $\Lambda \to \infty$
\cite{Gross-Politzer}.  
The renormalization of QCD does not involve an ultraviolet limit cycle.

We now turn to an independent theoretical development 
whose connection with RG limit cycles was first pointed out in
Ref.~\cite{Albeverio:zi}.
In 1970, Efimov discovered a remarkable effect 
that can occur in the 3-body sector for nonrelativistic particles 
with a resonant short-range 2-body interaction \cite{Efi71}.
If the particles are identical bosons and if the resonance 
is in the S-wave channel, the strength of the interaction
is governed by the S-wave scattering length $a$.
Efimov showed that if $|a|$ is much larger than 
the range $r_0$ of the interaction, there are shallow 3-body bound states
whose number increases logarithmically with $|a|/r_0$.
In the  resonant limit $a \to \pm \infty$, 
there are infinitely many shallow 3-body bound states 
with an accumulation point at the 3-body scattering threshold.
The ratio of the binding energies of successive states
rapidly approaches the universal constant
$\lambda_0^2 \approx 515$, where $\lambda_0 = e^{\pi/s_0} \approx 22.7$ 
and $s_0\approx 1.00624$ is a transcendental number.
In subsequent work, Efimov studied both the bound state spectrum and 
low-energy 3-body scattering observables for identical bosons 
with large scattering length \cite{Efi79}. He showed that the observables 
for different values of $a$ are related by a 
discrete scaling transformation in which $ a \to \lambda_0^n a$,
where $n$ is an integer, and lengths and energies are scaled by 
the appropriate powers of $\lambda_0^n$.

The Efimov effect can also occur for fermions with equal masses
and large scattering lengths if they have at least 3 
distinct spin states. 
Nucleons are examples of fermions with large scattering lengths.
The spin-singlet and  spin-triplet $np$ scattering lengths are
$a_s = -23.8$ fm and $a_t = 5.4$ fm.
They are both significantly larger than the $NN$ effective range,
which is $r_0 =1.8$ fm in the spin-triplet channel. Efimov used this 
observation  as the basis for a qualitative approach to the 
3-nucleon problem \cite{Efi81}.
His starting point was to take nucleons as point particles 
with zero-range potentials whose strengths are adjusted 
to reproduce the scattering lengths $a_s$ and $a_t$.
The effective range and higher order terms in the low-energy 
expansions of the phase shifts were treated as perturbations.
This approach works well in the 2-nucleon system,
giving accurate predictions for the deuteron binding energy.
This is no surprise; it simply reflects the well-known success 
of the effective range expansion in the 2-nucleon system \cite{Bet49}.
Remarkably, Efimov's program also works well in the 3-nucleon system
at momenta small compared to $m_\pi$. In the triton channel, 
the Efimov effect makes it necessary to impose a boundary condition 
on the wavefunction at short distances.  
The boundary condition can be fixed by using either 
the neutron-deuteron scattering length or the triton binding energy 
as input.  But if one of them is used as input, 
the other is predicted with surprising accuracy.
The accuracy can be improved by taking into account the effective range 
as a first-order perturbation \cite{Efi91}.

The Efimov effect was revisited by Bedaque, Hammer, 
and van Kolck within the framework of effective field theory (EFT)
\cite{Bedaque:1998kgkm}.
The problem of bosons with mass $m$ and large scattering length $a$ 
can be described by a nonrelativistic field theory with a 
complex-valued field $\psi$ and Hamiltonian density
\begin{eqnarray}
{\cal H} = (\hbar^2/2m) \nabla \psi^* \cdot \nabla \psi
+ g_2(\Lambda) (\psi^*\psi)^2.
\label{Heft}
\end{eqnarray}
For convenience, we set $\hbar=1$ in the following.
In the 2-body sector, the nonperturbative solution of the field theory 
can be obtained analytically.
Renormalization can be implemented by adjusting the 2-body
coupling constant $g_2(\Lambda)$ as a function of 
the ultraviolet cutoff $\Lambda$ so that the scattering length is $a$.
Other 2-body observables are then independent of $\Lambda$
and have the appropriate values for bosons with zero effective range.

In the 3-body sector, the nonperturbative solution of the field theory 
can be obtained by solving integral equations numerically.
These integral equations have unique solutions only 
in the presence of an ultraviolet cutoff $\Lambda$.  
The resulting predictions for 3-body observables,
although finite, depend on the ultraviolet cutoff and are
periodic functions of $\ln(\Lambda)$ with period $\pi/s_0$.
Bedaque, Hammer, and van Kolck showed that the quantum field theory 
could be fully renormalized to remove the residual dependence 
on $\Lambda$ in the 3-body sector by adding a 3-body interaction term
$g_3(\Lambda)(\psi^* \psi)^3$ to the Hamiltonian density in (\ref{Heft})
\cite{Bedaque:1998kgkm}.
The dependence of 3-body observables on the cutoff 
decreases like $1/\Lambda^2$ if the 3-body coupling constant 
has the form $g_3(\Lambda) = -4mg_2(\Lambda)^2 H(\Lambda)/\Lambda^2$, 
where the dimensionless function $H(\Lambda)$ has the form 
\begin{eqnarray}
H(\Lambda) = -{\sin[s_0 \ln(\Lambda/\Lambda_*) - \arctan(1/s_0)] 
	\over \sin[s_0 \ln(\Lambda/\Lambda_*) + \arctan(1/s_0)]} 
\label{Hlc}
\end{eqnarray}
for some value of $\Lambda_*$.
With this renormalization, 3-body observables have well-defined 
limits as $\Lambda \to \infty$, but they depend on 
the parameter $\Lambda_*$ introduced by dimensional transmutation.
Note that $H(\Lambda)$ is a periodic function of $\ln(\Lambda)$.
Thus the renormalization of the field theory involves 
an ultraviolet limit cycle.

An effective field theory (EFT) of nucleons 
with contact interactions only
has also been applied to the 3-nucleon problem 
\cite{Bedaque:1999ve,Hammer:2001gh}.
This contact EFT involves an isospin doublet $N$
of Pauli fields with two independent 2-body contact interactions:
$N^\dagger \sigma_i N^c N^{c \dagger} \sigma_i N$ 
and  $N^\dagger \tau_k N^c N^{c \dagger} \tau_k N$,
where $N^c = \sigma_2 \tau_2 N^*$.
Renormalization in the 2-body sector requires 
that the two coupling constants be adjusted as a function of the 
ultraviolet cutoff $\Lambda$ to obtain the correct spin-singlet
and spin-triplet scattering lengths $a_s$ and $a_t$.
Renormalization in the 3-body sector requires 
the 3-body contact interaction 
$N^\dagger \sigma_i N^c N^{c \dagger} \sigma_j N 
	N^\dagger \sigma_i \sigma_j N$
with a coefficient proportional to (\ref{Hlc}).
Thus the renormalization involves an ultraviolet limit cycle.

The low-energy few-nucleon problem can also be described 
by an EFT that includes explicit pion fields 
as well as contact interactions between the nucleons.  
The renormalization of such an EFT does not 
involve any RG limit cycle.  This suggests that the RG limit cycle 
in the EFT of Ref.~\cite{Bedaque:1999ve}
is just an artifact of the model.  We will argue to the contrary
that it is actually a hint of an infrared limit cycle in QCD.

Our argument is based on recent work in which an EFT 
with explicit pions was used to extrapolate nuclear forces 
to the chiral limit of QCD \cite{Beane:2002xf,Epelbaum:2002gb}.  
In this limit, the masses $m_u$ and $m_d$ of the up and down quarks 
are zero and the pion is an exactly massless Goldstone boson associated
with spontaneous breaking of the chiral symmetry of QCD.
According to the chiral extrapolation of Epelbaum, Mei{\ss}ner, 
and Gl{\"o}ckle \cite{Epelbaum:2002gb},
the small binding energy 2.2 MeV of the deuteron is a fortuitous consequence 
of the physical values of $m_u$ and $m_d$.
When extrapolated to the chiral limit, the deuteron has 
a much larger binding energy comparable to the scale 
$1/(m_N r_0^2) \approx 10$ MeV set by the $NN$ effective range.
Conversely, if extrapolated farther from the chiral limit,
the deuteron's binding energy decreases to 0 
and then it becomes unbound. 
\begin{figure}[tb]
\centerline{\includegraphics*[width=9cm,angle=0,clip=true]{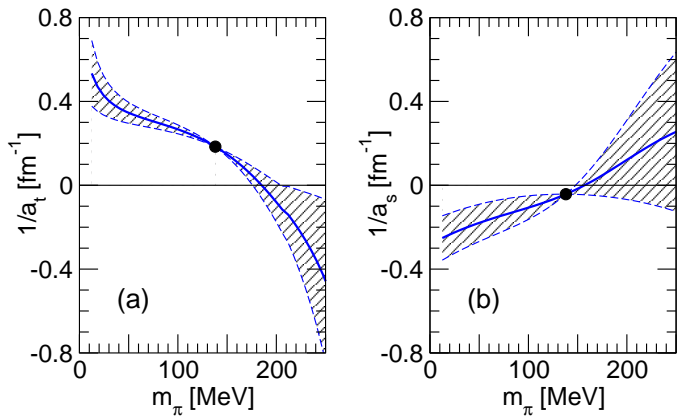}}
\caption{The inverse scattering lengths $1/a_t$ and $1/a_s$
as functions of $m_\pi$ as predicted 
by the EFT with pions of Ref.~\cite{Epelbaum:2002gb}.
}              
\label{fig:scat}
\end{figure}
This effect is illustrated in Fig.~\ref{fig:scat}, 
which shows the chiral extrapolation of the inverse scattering lengths 
$1/a_t$ and $1/a_s$ as functions of $m_\pi$ from Ref.~\cite{Epelbaum:2002gb}.
In the EFT of Ref.~\cite{Epelbaum:2002gb},
the coefficients of some of the 2-nucleon contact interactions are not 
well-constrained by the low-energy data.  The bands in Fig.~\ref{fig:scat} 
are obtained by varying those coefficients over natural ranges.
The width of the error band of course shrinks to zero 
at the physical value of $m_\pi$. 
The prediction of Ref.~\cite{Epelbaum:2002gb} is that 
the critical value $m_{\pi,t}$ at which $1/a_t=0$ 
is in the range 170 MeV $< \, m_{\pi,t} \, <$ 210 MeV,
which is not much larger than the physical value of $m_\pi$.
The extrapolation of $1/a_s$ has larger uncertainties.
It may increase to zero at some critical value
$m_{\pi,s}$ greater than 150 MeV, 
in which case the spin-singlet deuteron is bound 
for $m_\pi> m_{\pi,s}$, or $1/a_s$ may remain negative. 
Beane and Savage \cite{Beane:2002xf} have argued that the
extrapolation errors in the chiral limit are larger than estimated in 
Ref.~\cite{Epelbaum:2002gb}. For the small extrapolation to the 
region of larger $m_\pi$ where the deuteron becomes unbound, however, there 
is no controversy and the extrapolation errors are under control.

We now consider the chiral extrapolation of the 3-body spectrum. 
This could be calculated using an EFT 
with explicit pions.  Alternatively, the chiral extrapolation
can be calculated using the contact EFT of 
Ref.~\cite{Bedaque:1999ve}.
The inputs required are $a_s$, $a_t$, 
and $\Lambda_*$ as functions of $m_\pi$,
which can be calculated using an EFT with pions.
For the inverse scattering lengths $1/a_s$ and $1/a_t$,
we take the central values of the error
bands obtained from the chiral extrapolation in Ref.~\cite{Epelbaum:2002gb}.
The dependence of $\Lambda_*$ on $m_\pi$ could be determined 
from the chiral extrapolation of the triton binding energy
using an EFT with pions,
but this has not yet been calculated.
Like the inverse scattering lengths,
$\Lambda_*$ should vary smoothly with $m_\pi$.
For small extrapolations of $m_\pi$ from its physical value,
we can approximate $\Lambda_*$ by a constant.
We use the value $\Lambda_* =189$ MeV for $m_\pi=138$ MeV obtained
by taking the triton binding energy as the 3-body input.
\begin{figure}[tb]
\centerline{\includegraphics*[width=9cm,angle=0,clip=true]{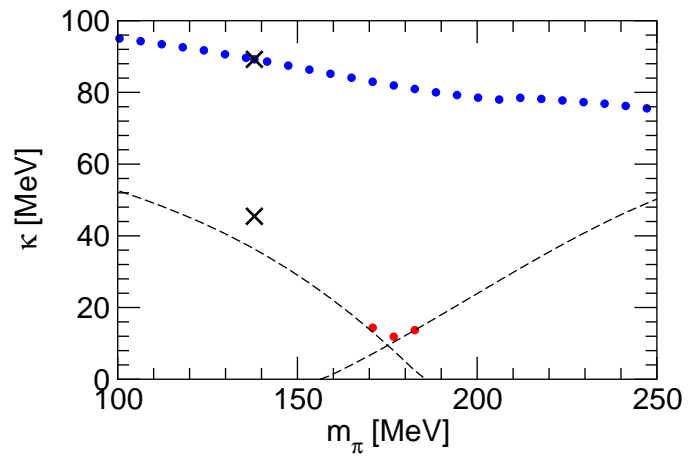}}
\caption{The binding momenta $\kappa=(mB_3)^{1/2}$ 
of $p n n$ bound states as a function of $m_\pi$ calculated
using the contact EFT of Ref.~\cite{Bedaque:1999ve}.
The circles indicate the triton ground state and excited state. 
The crosses give the binding energy of the physical
deuteron and triton, while the dashed lines give the thresholds
for decay into a nucleon plus a deuteron (left curve) or 
a spin-singlet deuteron (right curve).
}              
\label{fig:spec}
\end{figure}
In Fig.~\ref{fig:spec}, we show the 3-body spectrum in the triton 
channel as a function of $m_\pi$.  
The crosses give the binding momenta $\kappa=(mB_3)^{1/2}$ of the physical
deuteron and triton, while the dashed lines give the thresholds
for decay into a nucleon plus a deuteron (left curve) or 
a spin-singlet deuteron (right curve) in the large-$a$ approximation.
The circles indicate the triton ground state and excited state. 
In the region near $m_\pi \approx 175$ MeV where the decay threshold 
comes closest to the 3-nucleon threshold $\kappa = 0$,
the excited state of the triton appears in the spectrum.
The existence of this very shallow excited state is a hint 
that the system is close to an infrared limit cycle.

If, as in the case illustrated by Fig.~\ref{fig:spec}, 
the critical values $m_{\pi,t}$ and $m_{\pi,s}$ at which $1/a_t$ 
and $1/a_s$ go to zero satisfy $m_{\pi,s} < m_{\pi,t}$,
then the deuteron and spin-singlet deuteron are both bound
in the region $m_{\pi,s} < m_\pi <  m_{\pi,t}$.
Since the decay threshold never extends all the way down
to the 3-nucleon threshold $\kappa = 0$,
there cannot be an infrared limit cycle.  
We argue that it is possible, however, to tune the system to the 
critical RG trajectory for an infrared limit cycle 
by adjusting the up and down quark masses $m_u$ and $m_d$.
In the next-to-leading order chiral extrapolation of 
Ref.~\cite{Epelbaum:2002gb}, only quark mass dependent 
operators proportional to $m_u + m_d$ enter.
The extrapolation in $m_\pi$ can be interpreted as an 
extrapolation in the sum $m_u + m_d$,
with $m_u - m_d$ held fixed. Changing  $m_u - m_d$
changes the degree of isospin-symmetry breaking. 
Since the spin-singlet and spin-triplet channels correspond to 
isospin-triplet and isospin-singlet, respectively,
the scattering lengths $a_s$ and $a_t$ will respond differently 
to changes in $m_u - m_d$.  By tuning both $m_u$ and $m_d$,
it should be possible to make
$1/a_t$ and $1/a_s$ vanish simultaneously: $m_{\pi,t} = m_{\pi,s}$.
This point corresponds to a critical RG trajectory for an infrared 
limit cycle. At this critical point,
the triton has infinitely many increasingly-shallow excitations 
with an accumulation point at the 3-nucleon threshold.
As the excitations get more and more shallow, the ratio of the
binding energies of successive states rapidly approaches 
$\lambda_0^2 \approx 515$.  Only for the deepest states like the 
triton will there be significant deviations.
Now consider a renormalization group transformation 
that integrates out energies above some scale $\Lambda^2/m_N$.
As $\Lambda$ is decreased, the deepest 3-nucleon bound states are removed 
from the spectrum, leaving only those for which the deviations 
from the asymptotic ratio $\lambda_0^2$ are negligible.
Thus a limit cycle with a discrete-scaling-symmetry factor $\lambda_0$ 
is approached in the infrared limit $\Lambda \to 0$.

The error bands in Ref.~\cite{Epelbaum:2002gb} do not exclude
the critical values $m_{\pi,t}$ and $m_{\pi,s}$ from satisfying
$m_{\pi,t} < m_{\pi,s}$, in which case
the deuteron and spin-singlet deuteron are both unbound
in the region $m_{\pi,t} < m_\pi <  m_{\pi,s}$.
In this region, the decay threshold extends all the way down
to the 3-nucleon threshold $\kappa = 0$.
At either of the two critical values $m_{\pi,t}$ and $m_{\pi,s}$, 
there is a 2-nucleon bound state at threshold. 
However there is no Efimov effect, because the 3-body problem
is not sufficiently singular at short distances.
The exact infrared limit cycle can only be obtained by using an 
additional tuning parameter such as $m_u - m_d$
to set $m_{\pi,t} = m_{\pi,s}$.

The infrared RG limit cycle of QCD has important implications 
for attempts to derive nuclear physics from lattice gauge theory.
The computational effort for lattice simulations 
increases dramatically as the pion mass decreases 
and is prohibitive at the physical value.
Lattice simulations are typically carried out
at a value of $m_\pi$ that is 2-3 times larger than the 
physical value, and then a chiral extrapolation is made to
to $m_\pi = 138$ MeV.
Unfortunately, this requires extrapolating past the region
of $m_\pi$ where there is an RG limit cycle, 
which may introduce large extrapolation errors.
The proximity of physical QCD to the critical trajectory 
for the infrared limit cycle explains the success of 
Efimov's program \cite{Efi81} for describing the 3-nucleon problem using
zero-range forces or a contact EFT \cite{Bedaque:1999ve}. 
The apparent convergence of the effective-range corrections
for momenta of order $m_\pi$ \cite{Hammer:2001gh}
can be explained if the momentum expansion in the contact 
EFT is in powers of $p/m_\pi^*$, where $m_\pi^* \approx 175$ MeV
is the critical value of the pion mass, 
which is significantly larger than the physical value.
Our final conclusion is that the ultraviolet limit cycle 
in the contact EFT \cite{Bedaque:1999ve}
is not just an artifact of the model, but a hint of the 
existence of the infrared limit cycle of QCD.

This research was supported in part by DOE grant DE-FG02-91-ER4069. 
We thank U.-G.~Mei{\ss}ner for useful discussions, E.~Epelbaum
for providing the scattering lengths $a_s$ and $a_t$ as functions of
$m_\pi$, and R.J.~Furnstahl for comments on the manuscript.


\begin{thebibliography}{99}

\bibitem{Wilson:dy}
K.~G.~Wilson,
Rev.\ Mod.\ Phys.\  {\bf 55}, 583 (1983).

\bibitem{Wilson:1970ag}
K.~G.~Wilson,
	Phys.\ Rev.\ D {\bf 3}, 1818 (1971).

\bibitem{Glazek:2002hq}
S.~D.~Glazek and K.~G.~Wilson,
Phys.\ Rev.\ Lett.\ {\bf 89}, 230401 (2002).

\bibitem{Efi81}
V.~Efimov, Nucl.\ Phys.\ A {\bf 362}, 45 (1981).

\bibitem{Bedaque:1999ve}
P.~F.~Bedaque, H.-W.~Hammer, and U.~van Kolck,
	Nucl.\ Phys.\ A {\bf 676}, 357 (2000).

\bibitem{Gross-Politzer}
D.~J.~Gross and F.~Wilczek,
Phys.\ Rev.\ Lett.\  {\bf 30}, 1343 (1973);
H.~D.~Politzer,
Phys.\ Rev.\ Lett.\  {\bf 30}, 1346 (1973).

\bibitem{Albeverio:zi}
S.~Albeverio, R.~Hoegh-Krohn, and T.~T.~Wu,
Phys.\ Lett.\ A {\bf 83}, 105 (1981).

\bibitem{Efi71}
V.~N.~Efimov, Sov.\ J.\ Nucl.\ Phys.\ {\bf 12}, 589 (1971).

\bibitem{Efi79}
V.~N.~Efimov, Sov.\ J.\ Nucl.\ Phys.\ {\bf 29}, 546 (1979).

\bibitem{Bet49}H.~A.~Bethe, Phys.\ Rev.\ {\bf 76}, 38 (1949).

\bibitem {Efi91}
V.~Efimov, Phys.\ Rev.\ C {\bf 44}, 2303 (1991).

\bibitem{Bedaque:1998kgkm}
P.~F.~Bedaque, H.-W.~Hammer, and U.~van Kolck,
	Phys.\ Rev.\ Lett.\  {\bf 82}, 463 (1999);
	Nucl.\ Phys.\ A {\bf 646}, 444 (1999).

\bibitem{Hammer:2001gh}
H.-W.~Hammer and T.~Mehen,
	Phys.\ Lett.\ B {\bf 516}, 353 (2001);
P.~F.~Bedaque, G.~Rupak, H.~W.~Griesshammer, and H.-W.~Hammer,
Nucl.\ Phys.\ A {\bf 714}, 589 (2003).

\bibitem{Beane:2002xf}
S.~R.~Beane and M.~J.~Savage,
	Nucl.\ Phys.\ A {\bf 717}, 91 (2003); {\bf 713}, 148 (2003).

\bibitem{Epelbaum:2002gb}
E.~Epelbaum, U.-G.~Mei{\ss}ner, and W.~Gl{\"o}ckle,
	Nucl.\ Phys.\ A {\bf 714}, 535 (2003).

\end{thebibliography}
\end{document}